\documentclass[twocolumn,superscriptaddress,amsmath,amssymb,aps,prl,reprint,floatfix]{revtex4-1}

\usepackage{graphicx}
\usepackage{dcolumn}
\usepackage{bm}
\usepackage{float}

\begin{document}
\title{Time-resolved XUV Opacity Measurements of Warm-Dense Aluminium}

\author{S.M.~Vinko}
\email{sam.vinko@physics.ox.ac.uk}
\affiliation{Department of Physics, Clarendon Laboratory, University of Oxford, Parks Road, Oxford OX1 3PU, UK}

\author{V. Vozda}
\affiliation{Charles University, Faculty of Mathematics and Physics, Institute of Physics, Ke Karlovu 5, CZ-121 16, Prague 2, Czech Republic}
\affiliation{Institute of Physics, Czech Academy of Sciences, Na Slovance 2, Prague 8, Czech Republic}

\author{J. Andreasson}
\affiliation{ELI Beamlines, Institute of Physics, Czech Academy of Sciences, Na Slovance 2, CZ-182 21, Prague 8, Czech Republic}
\affiliation{Chalmers University of Technology, Department of Physics, G\"oteborg, Sweden}

\author{S. Bajt}
\affiliation{Deutsches Elektronen-Synchrotron DESY, Notkestrasse 85, Hamburg, Germany}

\author{J. Bielecki}
\affiliation{European XFEL GmbH, Holzkoppel 4, 22869 Schenefeld, Germany}

\author{T. Burian}
\affiliation{Institute of Physics, Czech Academy of Sciences, Na Slovance 2, Prague 8, Czech Republic}

\author{J. Chalupsky}
\affiliation{Institute of Physics, Czech Academy of Sciences, Na Slovance 2, Prague 8, Czech Republic}

\author{O. Ciricosta}
\affiliation{Department of Physics, Clarendon Laboratory, University of Oxford, Parks Road, Oxford OX1 3PU, UK}

\author{M. P. Desjarlais}
\affiliation{Pulsed Power Sciences Center, Sandia National Laboratories, Albuquerque, NM 87185, USA}

\author{H. Fleckenstein}
\affiliation{Center for Free-Electron Laser Science, DESY, Notkestrasse 85, 22607 Hamburg, Germany}

\author{J. Hajdu}
\affiliation{Laboratory of Molecular Biophysics, Department of Cell and Molecular Biology, Uppsala University, Husargatan 3, Box 596, SE-75124 Uppsala, Sweden}
\affiliation{ELI Beamlines, Institute of Physics, Czech Academy of Sciences, Na Slovance 2, CZ-182 21, Prague 8, Czech Republic}

\author{V. Hajkova}
\affiliation{Institute of Physics, Czech Academy of Sciences, Na Slovance 2, Prague 8, Czech Republic}

\author{P. Hollebon}
\affiliation{Department of Physics, Clarendon Laboratory, University of Oxford, Parks Road, Oxford OX1 3PU, UK}

\author{L. Juha}
\affiliation{Institute of Physics, Czech Academy of Sciences, Na Slovance 2, Prague 8, Czech Republic}

\author{M.F. Kasim}
\affiliation{Department of Physics, Clarendon Laboratory, University of Oxford, Parks Road, Oxford OX1 3PU, UK}

\author{E. E. McBride}
\affiliation{SLAC National Accelerator Laboratory, Menlo Park, CA 94025, USA}

\author{K. Muehlig}
\affiliation{Laboratory of Molecular Biophysics, Department of Cell and Molecular Biology, Uppsala University, Husargatan 3Box 596SE-751 24 Uppsala, Sweden}

\author{T.R. Preston}
\affiliation{European XFEL GmbH, Holzkoppel 4, 22869 Schenefeld, Germany}

\author{D. S. Rackstraw}
\affiliation{Department of Physics, Clarendon Laboratory, University of Oxford, Parks Road, Oxford OX1 3PU, UK}

\author{S. Roling}
\affiliation{Universit\"at M\"unster, Busso-Peus-Strasse 10, 48149 M\"unster, Germany}

\author{S. Toleikis}
\affiliation{Deutsches Elektronen-Synchrotron DESY, Notkestrasse 85, Hamburg, Germany}

\author{J. S. Wark}
\affiliation{Department of Physics, Clarendon Laboratory, University of Oxford, Parks Road, Oxford OX1 3PU, UK}
 
\author{H. Zacharias}
\affiliation{Universit\"at M\"unster, Busso-Peus-Strasse 10, 48149 M\"unster, Germany}

\date{\today}

\begin{abstract}
The free-free opacity in plasmas is fundamental to our understanding of energy transport in stellar interiors and for inertial confinement fusion research. However, theoretical predictions in the challenging dense plasma regime are conflicting and there is a dearth of accurate experimental data to allow for direct model validation. Here we present time-resolved transmission measurements in solid-density Al heated by an XUV free-electron laser. We use a novel functional optimization approach to extract the temperature-dependent absorption coefficient directly from an oversampled pool of single-shot measurements, and find a pronounced enhancement of the opacity as the plasma is heated to temperatures of order the Fermi energy. Plasma heating and opacity-enhancement is observed on ultrafast time scales, within the duration of the femtosecond XUV pulse. We attribute further rises in the opacity on ps timescales to melt and the formation of warm-dense matter.
\end{abstract}

\maketitle

The free-free opacity of a dense plasma at finite temperatures is a fundamental manybody problem on the boundary between plasma~\cite{Ron1963} and condensed matter physics~\cite{Sturm1973,Sturm1990}, with further practical applications across astrophysics, laser-plasma interactions and inertial confinement fusion research~\cite{Hu:2014}. At solid density, particle correlations, degeneracy and manybody effects all play an important role in determining how materials interact with light~\cite{Hopfield1965,Sturm1982}, invalidating the classical Coulomb-logarithm-based inverse bremsstrahlung picture widely applied in plasma physics modelling~\cite{Dawson:1962,Cauble:1985aa,Pfalzner:1998}. At the same time, the need to treat finite temperatures makes detailed approaches from condensed matter theory challenging to implement in practice, and increasingly unfeasible for temperatures exceeding of order 10~eV~\cite{Hollebon:2019}.

Because of its convenient electronic structure and the near-free behaviour of its valence electrons, aluminium irradiated at extreme ultra-violet (XUV) wavelengths is an ideal testbed to study free-free light-matter interactions at high electron densities. In this context, recent attempts have been made aimed at understanding historical discrepancies in the predicted and measured absorption coefficients in ground state Al at XUV wavelengths~\cite{Hollebon:2019}, with further investigations pushing well into the warm-dense matter regime~\cite{Kettle2016,Williams:2018}. Current theoretical predictions generally agree that the free-free absorption cross section should initially increase as the temperature of the system is raised to the Fermi temperature~\cite{Vinko2009a,Iglesias2010,Shaffer2017,Hollebon:2019}, before starting to fall off at higher temperatures according to the $T^{-3/2}$ temperature dependence of inverse bremsstrahlung (IB) theory~\cite{Decker:1994}. There are, however, considerable discrepancies in the predicted size, shape, extent and explanation of this effect.

Few experimental results of sufficient quality for opacity model benchmarking in regimes beyond the ground state are available in the literature. Kettle {\it et al.} conducted measurements of the XUV free-free opacity in laser-heated Al~\cite{Kettle2016}, but observed no significant change in the overall absorption at the 1~eV estimated temperature of the plasma. More recently, Williams {\it et al.} used 3~keV x-rays of the LCLS free-electron laser (FEL) to isochorically heat an Al foil which was subsequently probed by high-order laser harmonics~\cite{Williams:2018}. Average temperatures were estimated to have reached around 6~eV. The authors observed an increase in the absorption, an effect they attributed to a change in the ion structure, i.e., to melt and the formation of a warm-dense matter state. However, the measured transmission was averaged over a range of different temperatures and the authors did not report an absorption coefficient for the heated sample.

A common problem in opacity experiments is that it is challenging to measure accurately the plasma temperature, especially in the presence of strong gradients created by the heating source. If the opacity is a non-linear function of temperature it will also be increasingly difficult to interpret average measurements where the absorption takes place across a range of different plasma conditions. This makes measuring the temperature dependence of the absorption coefficient a formidable challenge, albeit an important one if theoretical approaches are to be quantitatively validated. Here we describe an alternative approach to tackling this issue. We combine a forward model for the self-heating of an FEL-irradiated sample with functional optimization and exploration algorithms to extract the temperature dependence of the absorption coefficient directly from an oversampled transmission dataset. This approach allows us to access the temperature dependence of the opacity without ever having to measure or average the temperature distribution. We then use these results to interpret time-resolved pump-probe measurements to infer the effect of electron and ion heating in the dense plasma regime.

\begin{figure}
\includegraphics[width=\linewidth]{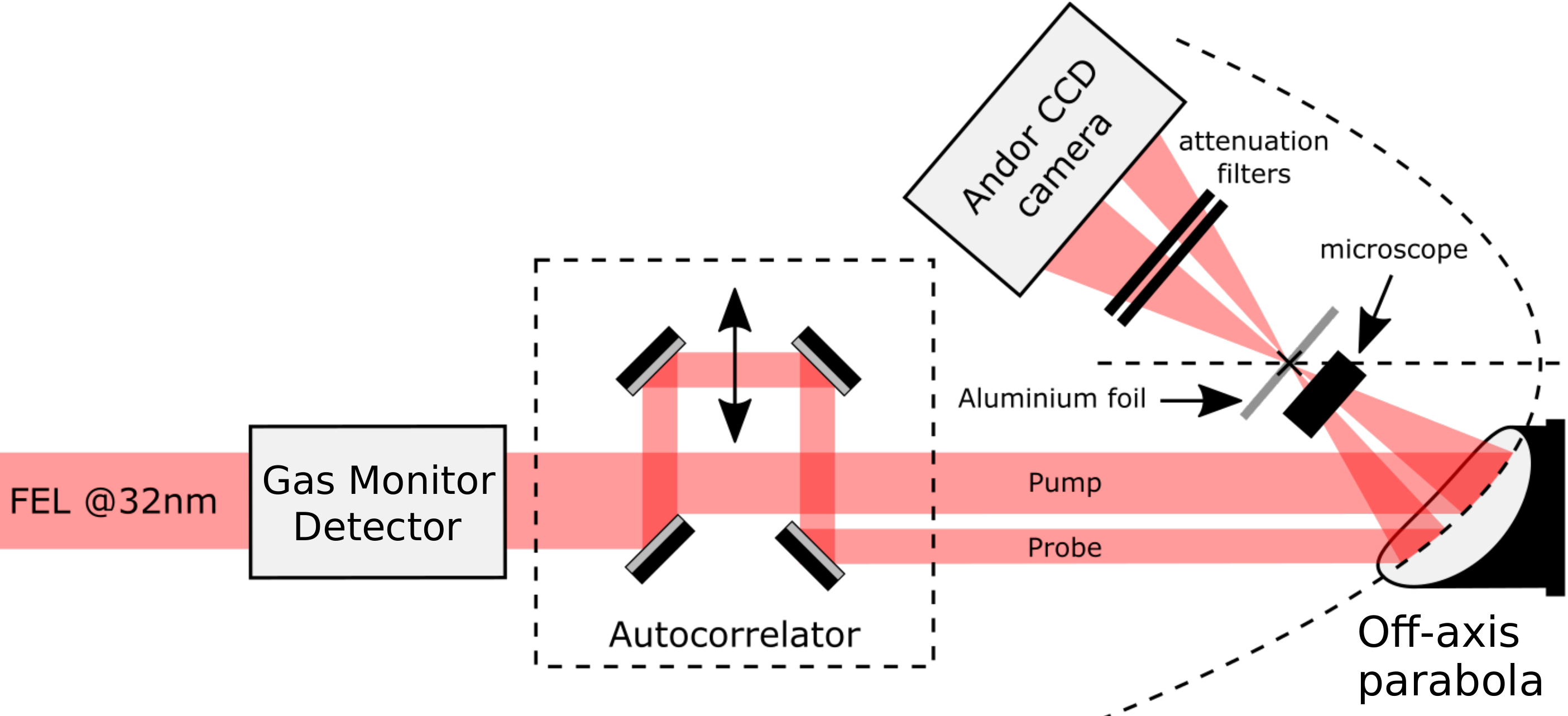}
\caption{Experimental setup: the XUV pulse is split and delayed in time before both pulses are focused on the sample by a multilayer coated off-axis parabolic mirror. The transmitted signal is measured by a downstream CCD camera.}
\label{FIG:setup}
\end{figure}

The experimental setup is illustrated schematically in Fig.~\ref{FIG:setup}. A 32~nm XUV pulse from the Hamburg FEL FLASH~\cite{Ackermann:2007} is split via an autocorrelator into a pump and probe pulse~\cite{Wostmann_2013}, with time-delays of up to 5 picoseconds. The pulses co-propagate toward a multilayer-coated off-axis parabolic mirror (OAP, reflectivity of 31\% at 32~nm) which focuses them onto thin Al foil targets. The pulses overlap spatially in focus only. Foil thicknesses of 200~nm and 300~nm were chosen to optimize changes of the transmitted signal as a function of heating. The FEL pulse duration was 100--150~fs, optimized to maximize the total energy in the beam. This pulse duration determines the overall time resolution of the experiment. Alignment of the OAP and microfocus characterization was done using ablative imprints on poly(methyl methacrylate)~\cite{Chalupsky:2010}. The spot sizes in focus were $(4.0 \pm 0.5)$ and $(5.4 \pm 0.7)$~$\mu$m$^{2}$ for the pump and probe, respectively. This is consistent with previous microfocusing efforts~\cite{Nelson:2009}. The focal spot and beam overlap are monitored via an on-axis, in-vacuum microscope, with a hole drilled through its optic to allow the FEL beam to pass through, while still providing micron-scale imaging resolution. The transmitted beam expands behind the target and illuminates a filtered CCD detector. The pulse energy was measured upstream by a gas monitor detector (GMD)~\cite{Tiedtke:2008} and was correlated with the observed signal intensity on the CCD in the absence of a target over a range of signal levels. This calibration is used to infer the energy of the pulse when a target is placed in focus and the transmission measured.

We irradiate samples at a photon energy of 38.8 eV (32~nm), significantly above the Al plasma frequency (15~eV) but below the first bound-edge of inner shell $2p$ states at 73~eV. The photons thus interact only with the near-free valence electrons, in the non-collective regime, making this an ideal prototypical system to study free-free absorption at electron densities exceeding $10^{23}$~cm$^{-3}$. The XUV pulse photoexcites the electrons which then collisionally redistribute their energy, and create a warm electron gas within a crystal ion lattice on time scales $<$100 fs~\cite{Vinko2010,Medvedev:2011,Dharmawardana:2015}. On picosecond time scales energy is transferred to the ions which heat and the system melts, forming a warm dense plasma~\cite{Cho:2011}.

While the absorption and electron heating processes are isochoric, significant gradients in temperature and electron density can be formed by the spatial distribution of the XUV pulse both on the surface of the sample and volumetrically. The sample is therefore not homogeneously heated, and a single, average temperature is a poor descriptor of a plasma for determining free-free opacities.

The presence of gradients in FEL-irradiated systems thus seems to be a major disadvantage to plasma investigations. However, because gradients are related to the imprint of the XUV intensity distribution in focus on target, they are both predictable and measurable. Unlike in optical laser-plasma experiments, here there is no mechanism for MeV hot-electron generation, the Keldysh parameter is negligible, and the mean free path of the electrons created by the XUV pulse is of order of 10~nm, small compared with the FEL spot size of a few micron. The intensity distribution can be complex, determined by source profile effects, beam quality and beamline and focusing optics, but it remains fixed over the course of the experiment. Crucially, it can be measured via ablative imprints, and the 3D structure of the pulse can be reconstructed~\cite{Chalupsky:2013}. This provides a fascinating opportunity not only to dispose of the problems generated by gradients in the first place, but to use them to our advantage in understanding the absorption process itself.

\begin{figure}
\includegraphics[width=\linewidth]{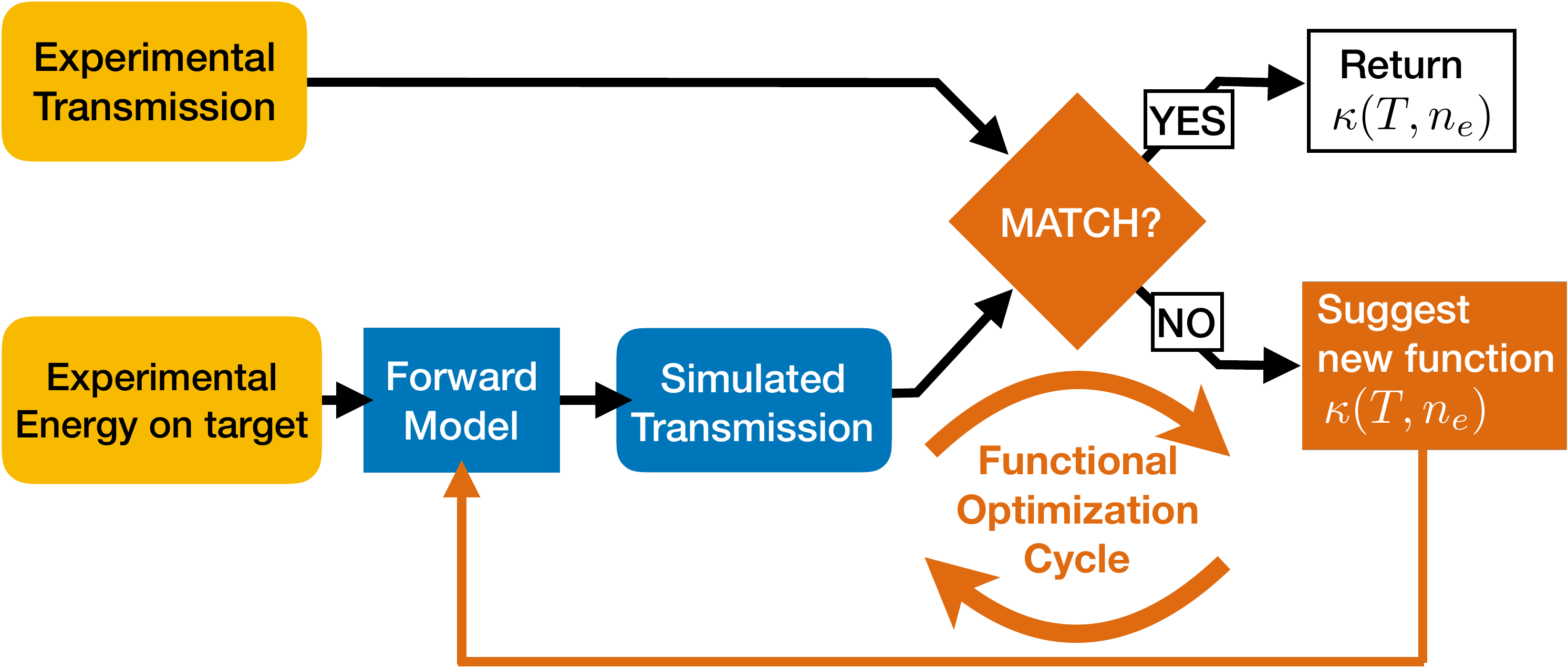}
\caption{Schematic representation of the functional optimization approach used to extract the opacity as a function of electron temperature and density by constraining it to best-match the dataset of integrated transmission measurements.}
\label{FIG:MachineDiscovery}
\end{figure}

We proceed as follows. Firstly, we note that the XUV intensities here are relatively low, far below the non-linear regime threshold of $\sim10^{16}$~Wcm$^{-2}$~\cite{Nagler2009}, so photoabsorption is a linear process and follows the Beer-Lambert law:
\begin{equation} \label{Beer-Lambert}
\frac{\mathrm{d} I}{\mathrm{d} x} = -\kappa(T_e, n_e) x,
\end{equation}
with $I$ the pulse intensity, $x$ the depth into the target and $\kappa(\cdot)$ the absorption coefficient, a function of the free-electron temperature $T_e$ and density $n_e$. As the energy is absorbed by the electrons, knowledge of the electron equation of state (EoS) and of the functional form of the absorption coefficient thus fully determines the heating dynamics of the irradiated sample. 

\begin{figure*}
\includegraphics[width=\linewidth]{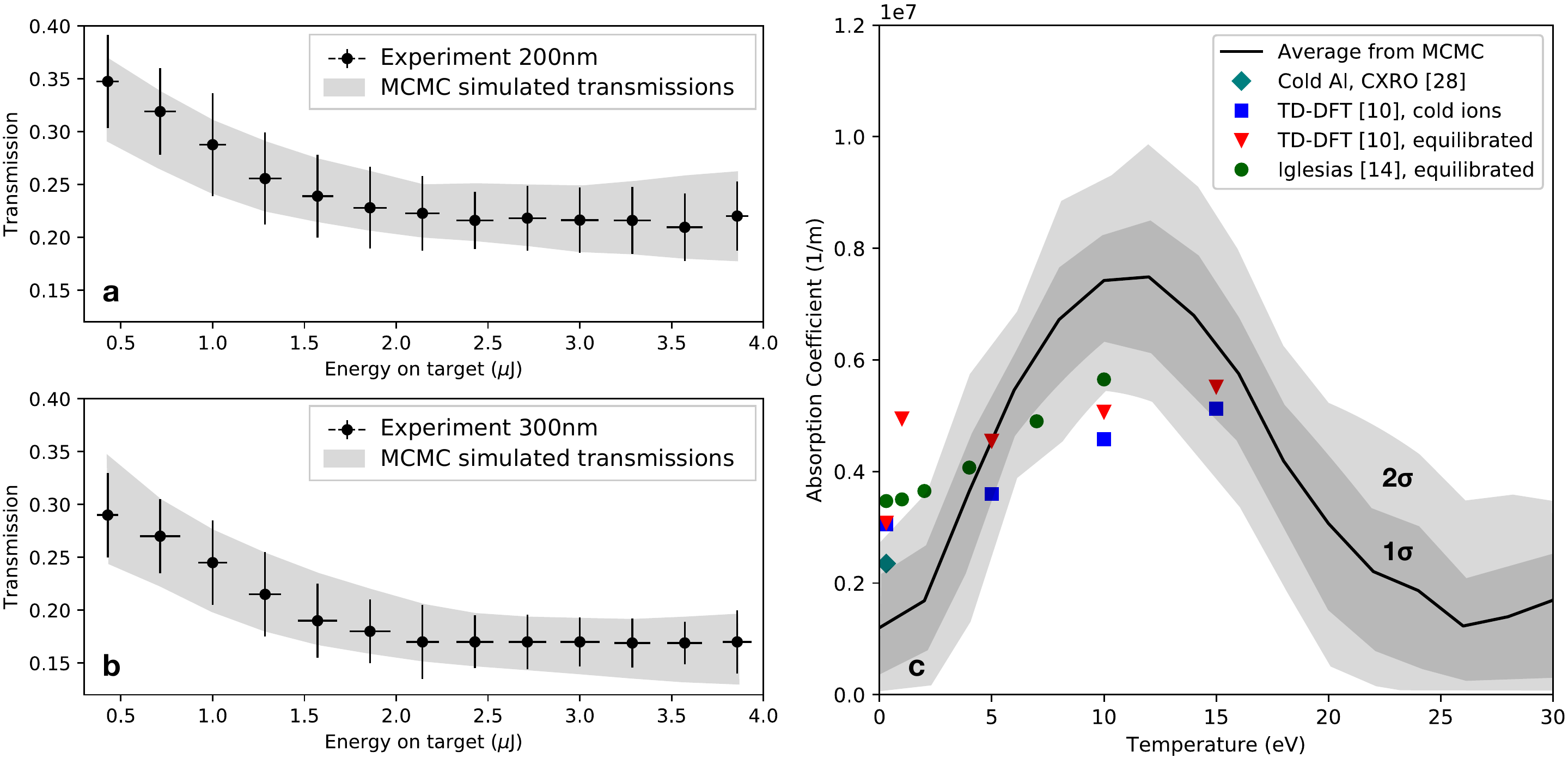}
\caption{Total experimental and simulated transmission measured through 200~nm (a) and 300~nm (b) foils for a range of different total FEL energies on target, alongside the simulated transmissions within a 2$\sigma$ band. c) Absorption coefficient extracted by our functional optimization approach. The curve is constrained by the data up to the peak temperatures of 26~eV. Bands are determined by MCMC calculations.
Also plotted are the theoretical predictions based on ref.~\cite{Hollebon:2019} assuming either cold ions or an equilibrated system, the IB-based theory of Iglesias~\cite{Iglesias2010}, and the cold absorption from the CXRO database~\cite{CXRO}.}
\label{FIG:OpacityCurve}
\end{figure*}

A simple EoS can be constructed by assuming the electronic structure of Al is well-described by finite-temperature density functional theory (DFT)~\cite{Mazevet2008,Vinko2010,Williams:2018}. The energy density of the electrons at a temperature $T_e$ is given by
\begin{equation} \label{EnergyDensity}
E/V = \int_0^{\infty} \varepsilon \; \mathcal{D}(\varepsilon) f_{\rm FD} (\varepsilon, T_e; \mu) \;  \mathrm{d} \varepsilon,
\end{equation}
where we set the energy of the bottom of the valence band to 0, $\mathcal{D}(\varepsilon)$ denotes the density of states, $f_{\rm FD} (\cdot)$ the Fermi-Dirac distribution and $\mu$ the chemical potential. To calculate the density of states across the temperatures of interest we used the ABINIT code~\cite{Torrent2008,Gonze2009,Gonze2016a}. We observe that the valence electrons remain at constant (solid) density up to temperatures around 10~eV, but above that thermal ionization of the $2p$ and $2s$ states produces a temperature-dependent valence electron density $n_e(T_e)$. This process is fully accounted for in the DFT modelling. We will assume that electrons excited by 30-40~eV above the Fermi energy can thermalise most of their energy on timescales short compared with the duration of the XUV pulse. This assumption will be validated later by our experimental results. 
Self-consistently solving Eqs.(\ref{Beer-Lambert}) and (\ref{EnergyDensity}), given some form for $\kappa(T_e, n_e)$, and using the XUV energy distribution on target, constitutes a forward model for XUV self-heating. According to this model we can calculate the energy absorption and target heating as the XUV pulse propagates through our sample. By the end of the pulse we obtain a 3D map of the conditions present, and a measure of the total transmitted pulse energy.

Given that we measure the energy distribution on target, one can hope to extract the absorption coefficient as a function of temperature and density purely from measurements of the total energy in the beam before and after interacting with the sample. Clearly, each such datapoint systematically encodes a range of different plasma conditions within the absorbing system, and so the gradients contain all the information on the energy-density dependence of the absorption coefficient. Thus, by sampling the transmission of many pulses with different total energy content, we can reconstruct the functional form of the absorption coefficient. What is perhaps surprising in this approach, illustrated in Fig.~\ref{FIG:MachineDiscovery}, is that it allows us to extract the absorption coefficient as a function of temperature without ever having to explicitly measure the temperature.

The experimental transmission measured through 200 and 300~nm foils for a range of different FEL energies on target is displayed in Figs.~\ref{FIG:OpacityCurve}a and~\ref{FIG:OpacityCurve}b, for a total of over 5000 single shots split into 13 equally-spaced energy bins. The uncertainties represent the 1-$\sigma$ scatter within each bin. Our forward model is run for each point shown, and the optimization objective is to find a form for $\kappa(T_e, n_e)$ that produces the best match to every point in the dataset. The absorption function is sampled every 2~eV on a temperature grid up to 50 eV. The modelling suggests peak temperatures of 26~eV are reached at the highest experimental pulse energies.

Uncertainties on the data remain significant as it is challenging to measure the transmission to better than 10-20\%. As such, a calculation of how these uncertainties impact the final form of $\kappa(T_e,n_e)$ is needed. For this we use Bayesian inference to explore the space of functions able to represent the absorption coefficient, via a combination of optimization and Markov Chain Monte Carlo (MCMC) algorithms~\cite{Kasim:2019}. We start by finding the best-fit solution using the stochastic CMA-ES optimization algorithm~\cite{cmaes}, and use it as a starting point for the ensemble MCMC~\cite{Goodman:2010} using only the stretch move.
For the sampling we deployed 32 walkers in parallel collecting a total of over 300,000 samples after 100,000 were discarded for burn-in. The resulting absorption coefficient is shown in Fig.~\ref{FIG:OpacityCurve}c, with 1$\sigma$ (68\%) and 2$\sigma$ (95\%) confidence intervals.
We also show the theoretical predictions from time-dependent DFT calculations based on the work of Hollebon {\it et al.} for both an equilibrated system~\cite{Hollebon:2019} and for a system with ions at 300~K, and from Iglesias~\cite{Iglesias2010}.

\begin{figure*}
\includegraphics[width=\linewidth]{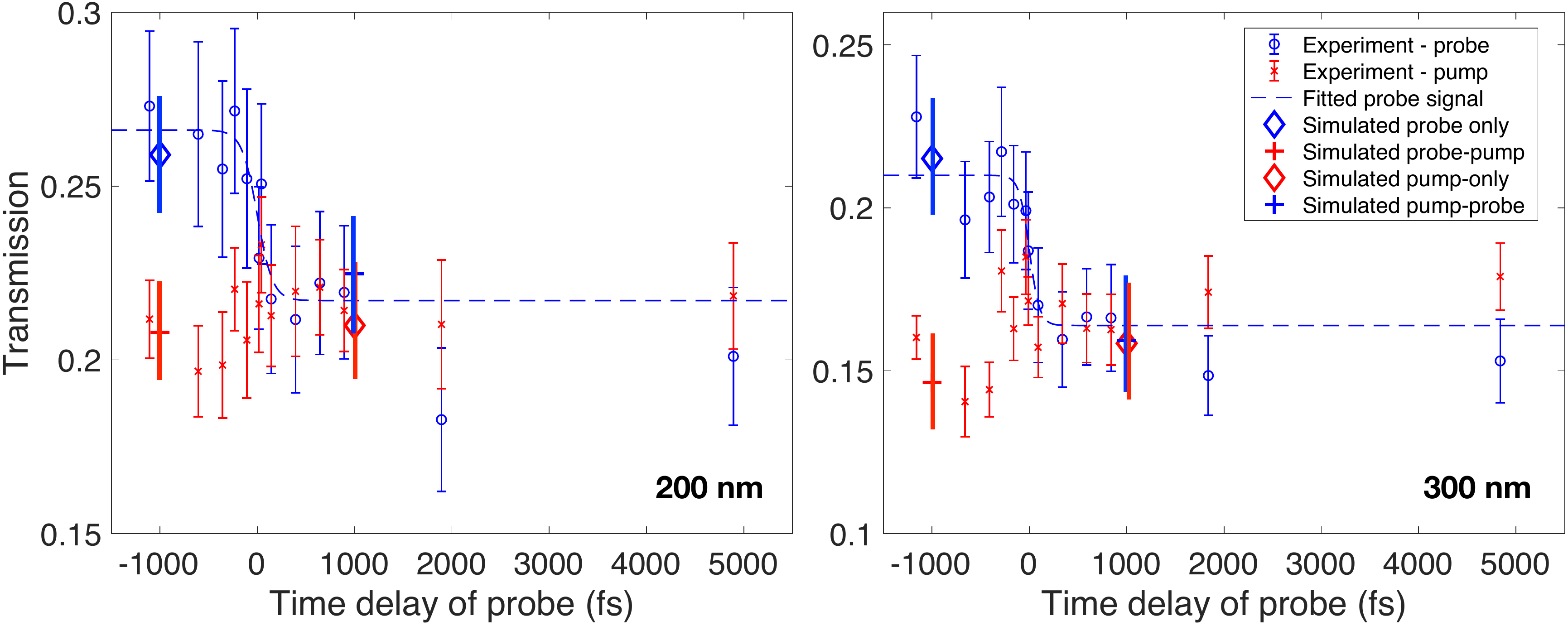}
\caption{Pump-probe transmission measurements in Al samples 200 and 300~nm thick. Negative times correspond to the probe pulse arriving first and positive times to the pump arriving first. The simulated transmissions for a single (diamond) or two pulses (cross) are also shown, where the forward model was used with the absorption coefficient extracted from the single-short self-heating data.}
\label{FIG:PumpProbe}
\end{figure*}

Having understood how a single XUV pulse heats the sample, we now turn our attention to the pump-probe measurements. The probe pulses typically have an energy of around 1--1.5~$\mu J$, while the pump varies between 2--4~$\mu J$. We see from Fig.~\ref{FIG:OpacityCurve} that a 1~$\mu J$ pulse already gives rise to heating and a change in the transmission from the cold value. The heating of the probe is thus not negligible, but experimentally is was not possible to further reduce its energy and still acquire reliable transmission values.
The measured pump-probe transmission is shown in Fig.~\ref{FIG:PumpProbe} for both 200 and 300~nm foils. Negative times indicate the prior arrival of the probe pulse. We observe a marked decrease in the transmission of the probe pulse around $t=0$, the size of which is consistent with the absorption coefficient extracted from single-shot measurements. At negative times the probe pulse arrives first and mildly heats the system, but as we move to the zero-delay region, where both pump and probe pulses hit the target at the same time, more heating occurs leading to higher absorption. At later times the system probed by the second pulse remains heated on picosecond timescales so the transmission remains suppressed. We fit the probe data at $\pm1$~ps with a sigmoid, illustrating that the change in the absorption takes place on femtosecond timescales, within the duration of the pulse. This supports our initial assumption of thermalization within the XUV pulse in our heating model. The pump pulse shows a similar behaviour to the probe, but because it contains more energy the change in transmission at $t=0$ is smaller. In contrast to the probe pulse, the pump arrives earlier in time for positive delays, so here the transmission is higher as the total heating is lower. For negative times the pump comes after the probe, and so the transmission is further suppressed. The width of the transition here is harder to estimate but is consistent with the timescales observed for the probe pulse.

We ran our forward model on the pump-probe data using the absorption coefficient extracted from the self-heating results. We show these results with the diamond and cross symbols in Fig.~\ref{FIG:PumpProbe}: the diamonds indicate the simulation of a single pulse only, while the crosses indicate the transmission of a pulse through an already heated sample. We see broad agreement with the experimental data for all cases. The probe pulse is slightly larger than the pump, and the outer wings of the pulses do not overlap. However, this does not seem to matter much from the modelling, due to the small values of the absorption coefficient at low temperatures. We note that the values of the simulated transmissions assume a single well-defined energy on target, but experimentally data was collected over many shots with a considerable spread in energy, in part determined by the FWHM of the reflectivity peak of the OAP. Here we show the mean of the data as an indicative value, with the error bar determined by the variation in the data.

Theoretically there are two processes that give rise to an enhanced absorption at finite temperatures. The first is electronic, due to thermal broadening of the plasmon peak~\cite{Vinko2009a}, increases in the many-body screening length and a reduction in the electron degeneracy as the electrons heat~\cite{Iglesias2010}. The second is due to the change in the ion structure factor as the ions heat and the system melts~\cite{Vinko2009a,Hollebon:2019}. The electron contribution is readily observed on femtosecond timescales in our data. In contrast, the effects of melt are expected to take place over longer, picosecond timescales. From the data in Fig.~\ref{FIG:PumpProbe} we observe an indication of an additional systematic decrease in the probe transmission for time delays beyond 1~ps. Our results contain only four points in this region so the effect is challenging to quantify accurately, but the change in the transmission implies a further 10-20\% enhancement of the absorption coefficient due to the formation of warm dense matter.

In summary, we presented time-resolved measurements of the free-free opacity in XUV-heated Al. By using a functional optimization approach to the interpretation of the experimental transmission data we were able to use the gradients created by the FEL isochoric heating process to extract the absorption coefficient as a function of electron temperature and density, without needing to measure the electron temperature, density or ionization explicitly. We find that the absorption increases initially with heating and peaks at temperatures around the Fermi temperature. While the heating of the electron subsystem dominates the overall change in the opacity, both electronic and ionic effects lead to an increased absorption at finite temperatures. We find hot opacities that are significantly larger than predicted by theoretical calculations.

\begin{acknowledgments}

Portions of this research were carried out at the FLASH facility. We are grateful to the scientific and technical staff of FLASH for their outstanding facility and support.
This work was supported by the European Community's Seventh Framework Programme (FP7/2007-2013) under the grant agreement CALIPSO 312284 (EU Support of Access to Synchrotrons/FELs in Europe ).
S.M.V. is a University Research Fellow of the Royal Society. S.M.V., M.F.K, and J.S.W. acknowledge support from the U.K. EPSRC grant EP/P015794/1 and from the Royal Society.
T.B., J.Ch., V. H., L.J. and V.V. appreciate support by CMEYS (grants LTT17015 and CZ.02.1.01/0.0/0.0/16\_013/0001552) and CSF (grants 17-05167S a 19-03314S).
J.A. and J.H. are supported by projects CZ.02.1.01/0.0/0.0/16\_019/0000789 (ADONIS) and CZ.02.1.01/0.0/0.0/15\_003/0000447 (ELIBIO) from European Regional Development Fund. J.A. acknowledges support from the R\"ontgen {\r A}ngstr\"om Cluster, Chalmers Area of Advance; Material Science and the Ministry of Education, Youth and Sports as part of targeted support from the National Programme of Sustainability II.
S.R and H.Z. acknowledge support from the BMBF under grant 05K13PM2.

\end{acknowledgments}


%\bibliographystyle{apsrev4-1}
%\bibliography{MyCollection}

\end{document}